# Weak antilocalization effect in LPE-grown p-Hg$_{0.8}$Cd$_{0.2}$Te thin film and the evidence of Te-precipitation

R.Yang, L.M.Wei and G.L.Yu

*National Laboratory for Infrared Physics, Shanghai Institute of Technical Physics, Chinese Academy of Science, Shanghai 200083, People's Republic of China*

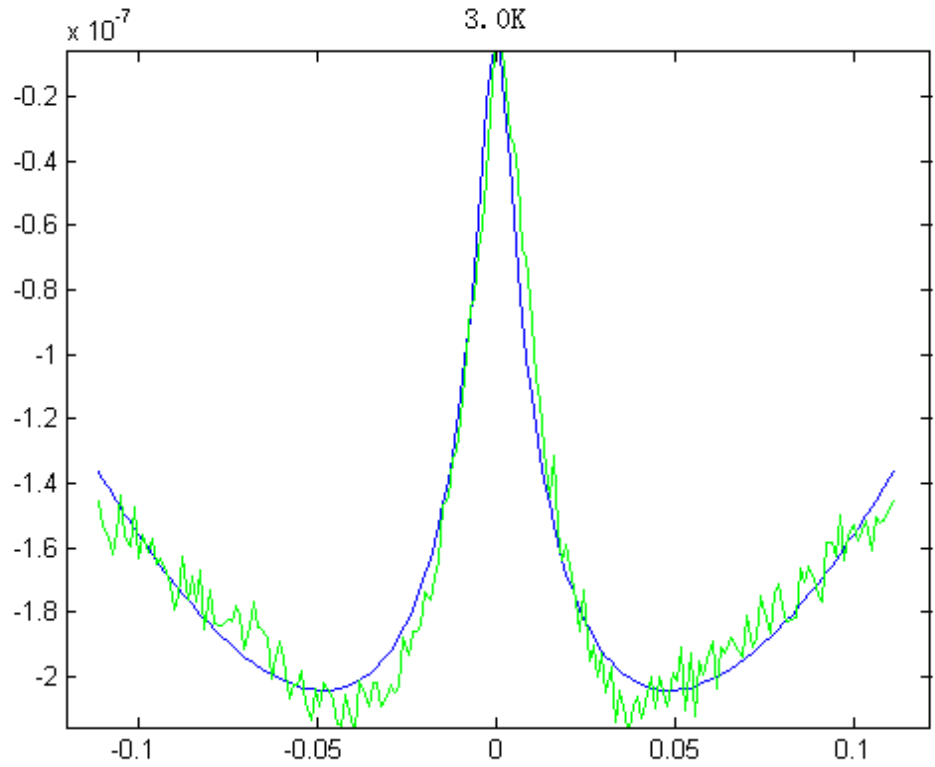

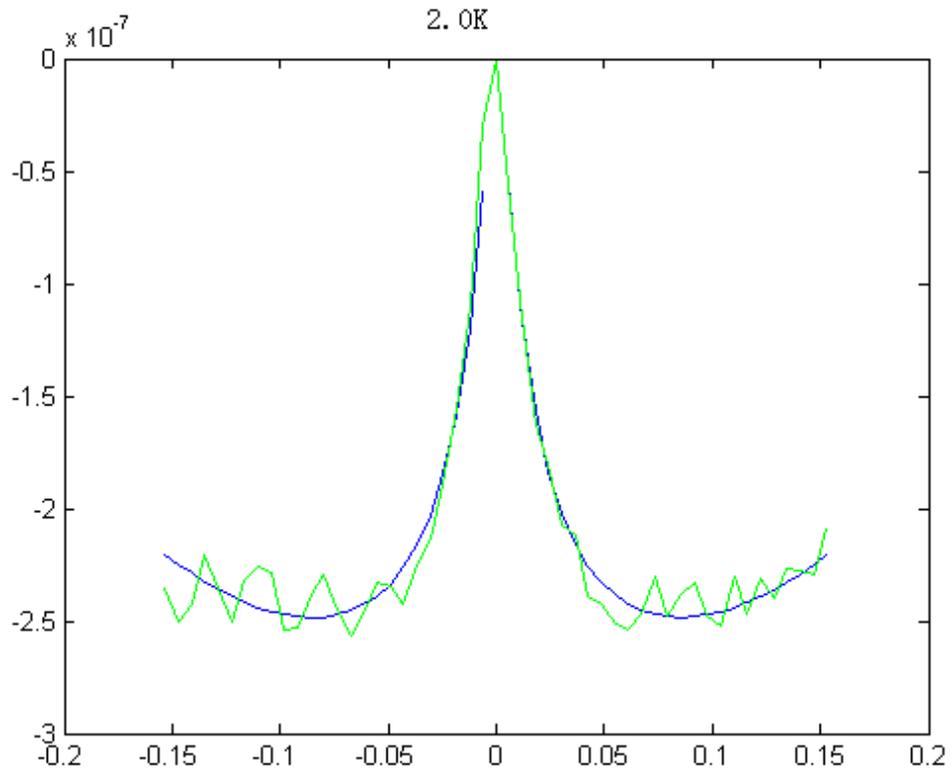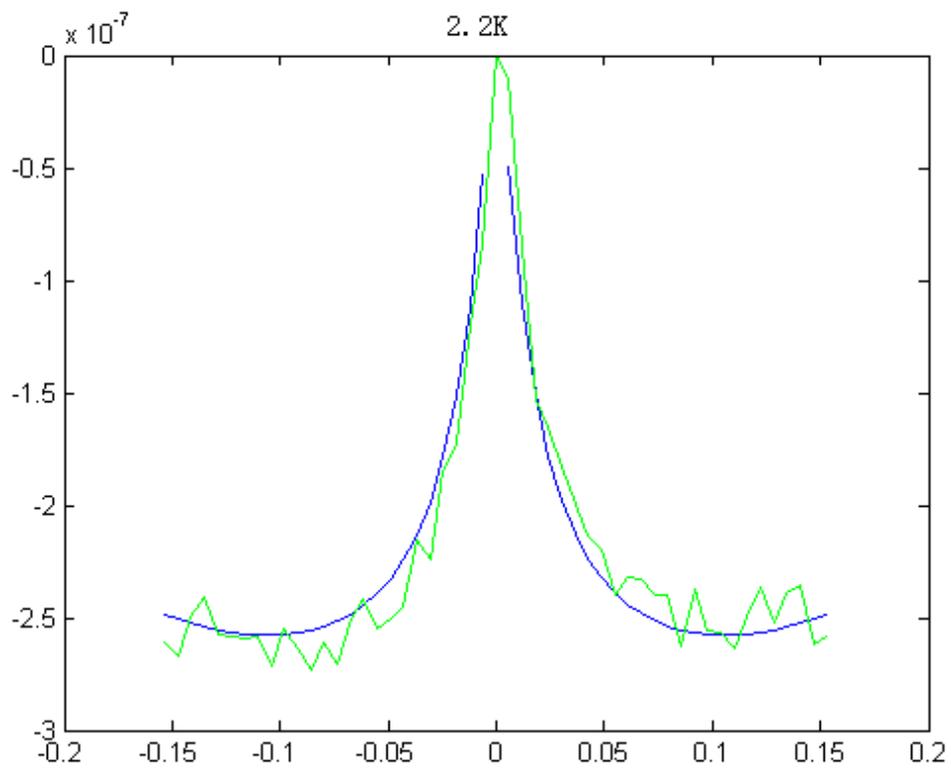

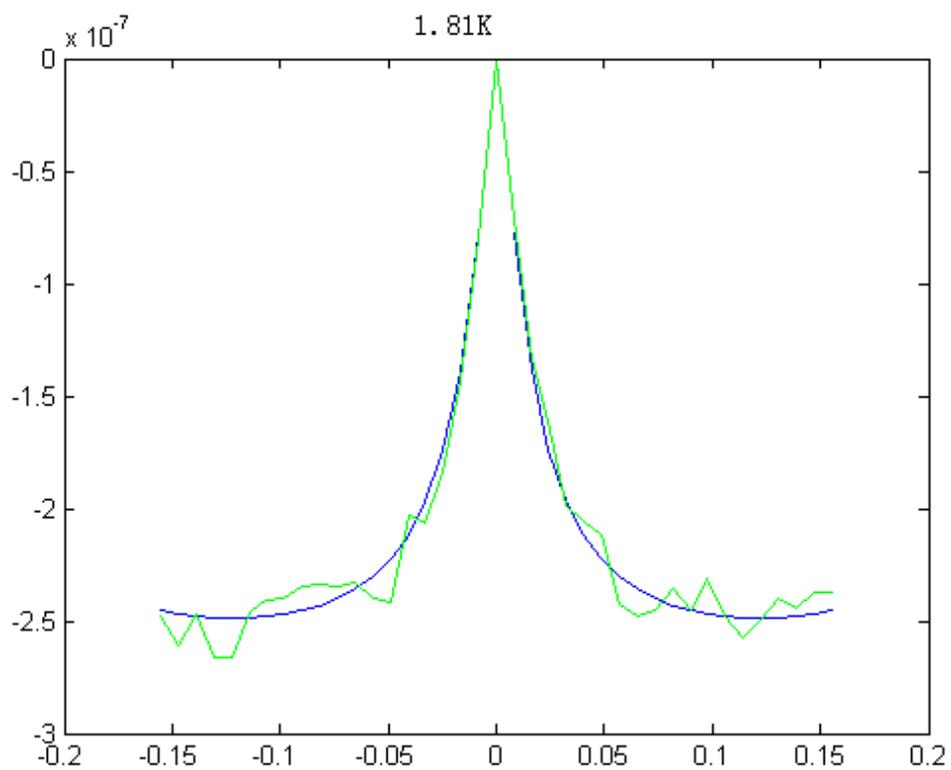

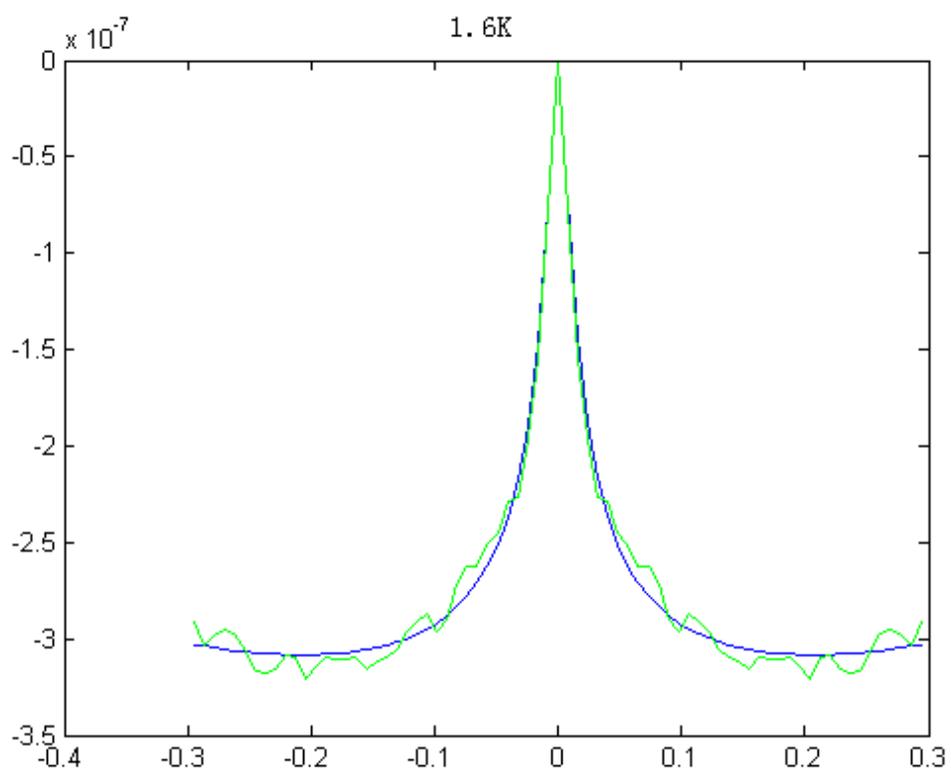

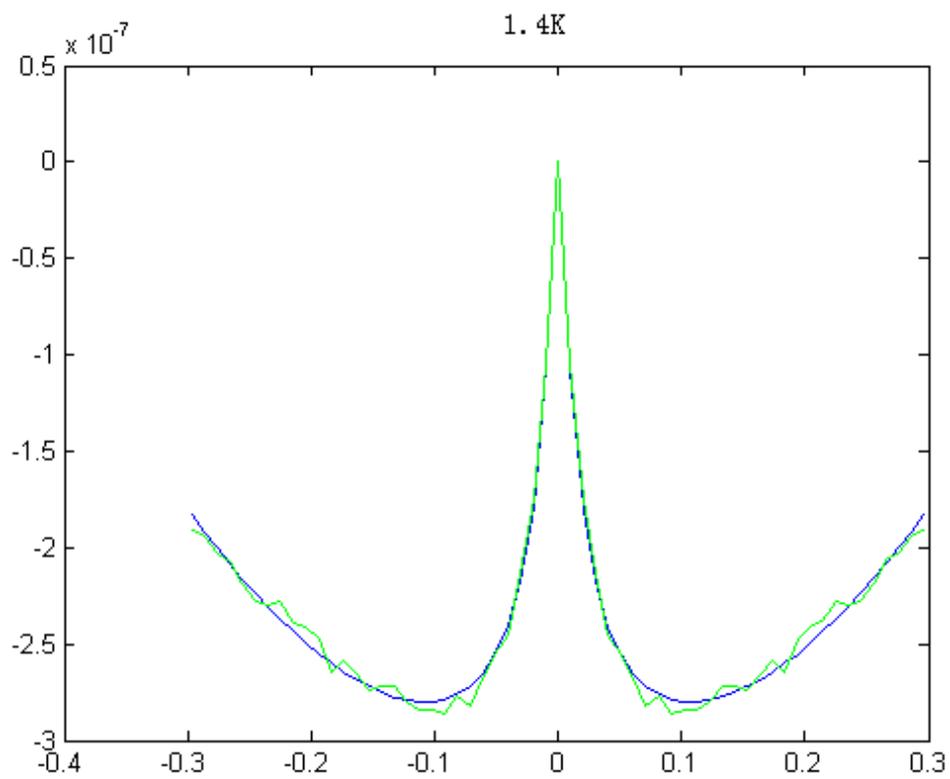

FIG.1

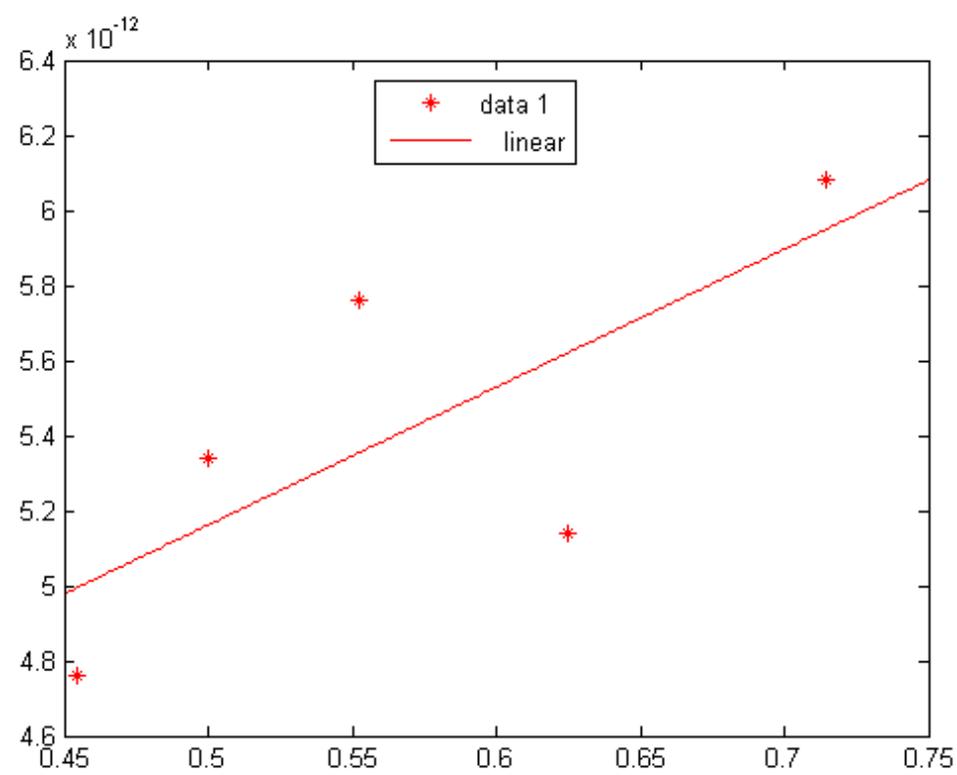

FIG. 2 Plot of taophi~1/T

The weak antilocalization effect is observed in a p-type $Hg_{0.8}Cd_{0.2}Te$ thin film with thickness ~10 micrometers.

In order to nail down the origin of the antilocalization effect, a detail investigation of the composition and carrier species is necessary.

The MCT thin film is grown by LPE method; there are many precipitations and inclusions in this kind of sample. According to previous research, in LPE grown MCT sample, there are Te-inclusion and other precipitations existing in MCT matrix. Moreover, on its surface, there usually is a Te-rich layer in which Te-inclusion, $CdTeO_2$-inclusion and other type of precipitations can be found[1,2,3].

The carrier species of LPE grown MCT sample is usually complex. We also perform preliminary multi-carrier fitting about our data. The result shows that there are more than one carrier species existing in our sample. Especially, the analysis implicates that there is a surface carrier species. This result is consistent with previous research concerning p-HgCdTe[4].

After investigation of sample composition and carrier species, we analyze the weak antilocalization effect observed in our MCT thin film. Based on the assumption that one species dominates conducting in our sample. I try to fit the data with varies editions of widely-used weak antilocalization models which don't take intervalley-scattering and isospin that are closely connected to electronic structure of specific material into consideration, namely, two editions of 2D model (HLN model and ILP model); two editions of quasi-2D model (replacing diffusion coefficient in previous 2D model with bulk value); two editions of 3D antilocalization model (Kawabata model and Fukuyama-Hoshino model). All of these attempts fail to get an acceptable fitting result. Similar weird antilocalization effect has also been observed in other groups' research[7,8].

I also apply a 2D model which deals with the weak antilocalization effect of holes in Te crystal and takes the specific intervalley-scattering and isospin that are connected to Te's unique Brillouin zone into consideration[5]. The fitting of experimental data with this model is excellent, see FIG.1. So, here comes a possible explanation concerning the origin of weak antilocalization effect in our HgCdTe sample that the weak antilocalization is caused by Te-inclusion. However, in the temperature where weak antilocalization emerges, the conduction type seems to be n type for our sample, it's a confusing fact that seems to be in consistent with Te-inclusion explanation. A possible explanation is that weak antilocalization formula of hole in Te can be applied to electrons as well. In addition, the 2D nature of the weak antilocalization effect is supported by multi-carrier analysis, thus the application of 2D model is also valid. The taophi~1/T plot is approximately linear and is consistent with the relation usually followed by taophi~1/T in weak antilocalization effect.

Aside from Te-precipitation explanation, another possible explanation is that the weak antilocalization effect is caused by the existence of Te-like electronic structure in inversion layer on p-HgCdTe that can lead to similar intervalley scattering and similar weak antilocalization mechanism in which intervalley scattering plays an important role. At the first glance, it's plausible because CdTeO$_2$ precipitation has the same electronic structure as that of Te[6].

Furthermore, besides these two explanations of extrinsic origin of weak antilocalization effect in HgCdTe, intrinsic scenario is also possible. It's possible that the weak antilocalization effect is caused by intrinsic properties of HgCdTe such as a weak antilocalization mechanism concerning intervalley scattering of HgCdTe, and the resulting formula has the same form as that in Te.

From above analysis, the most plausible explanation of the weird weak antilocalization effect observed in our p-HgCdTe sample is that the observed weak antilocalization effect arises from Te-inclusion especially Te-inclusion in the surface layer of our LPE-grown MCT sample. The effect is not intrinsic to MCT matrix. More analysis like infrared spectrum will help to reveal the existence and concentration of Te-inclusion in MCT sample.

From above analysis, we can see that there are some obstacles impeding us from the observation of intrinsic weak antilocalization effect in MCT film. One obstacle is the existence of precipitations, like Te-inclusion. This may provide an explanation for previously observed weird weak antilocalization effect that can't be explained within conventional models[7][8]. One way to handle this problem is to use better method like MBE method to grow sample. For Te-inclusion, using n-type MCT for investigation can avoid the interference of Te-inclusion upon data explanation. Another obstacle is that MCT sample grown by LPE is tend to be in the strong localization regime rather than weak localization regime under low-temperature. Using better synthesis methods or Cd-content and proper doping may help to overcome this obstacle.

In order to observe weak antilocalization effect in Hg-based narrow-band semiconductors, there are two proposals. One proposal is to use MBE-grown HgTe thin film, because clear weak antilocalization effect that can be explained quite well within conventional models has been observed in previous researches[9].
Another proposal is to use As-doped MBE-grown HgCdTe thin film.